\documentclass[twocolumn,aps,prb,twocolum,superscriptaddress,bibnotes,amsmath,amssymb,floatfix,footinbib,longbibliography]{revtex4-2}
\usepackage[markup=blue, authormarkupposition=left]{changes}
\usepackage{soul}
\usepackage[utf8]{inputenc}
\usepackage[english]{babel}
\usepackage{amsmath,amsfonts,amssymb}
\usepackage[T1]{fontenc}
\usepackage{url}
\usepackage{changes}
\usepackage{soul}
\usepackage{amsmath}
\usepackage{amsfonts}
\usepackage{amssymb}
\usepackage{xcolor}
\usepackage{siunitx}
\usepackage{epstopdf}
\usepackage{graphicx}
\usepackage{changes}
\usepackage{float}
\usepackage[labelfont=bf, justification=centerlast]{caption}

\graphicspath{{./Figures/}}
\usepackage[
bookmarks=true,
colorlinks=true,
linkcolor=blue,
urlcolor=blue,
citecolor=blue,
pdftex,
bookmarks=true,
linktocpage=true, % makes the page number as hyperlink in table of content
hyperindex=true
]{hyperref}

\newcommand{\SiN}[0]{Si$_3$N$_4$~}

\begin{document}
	
	\title{Hertz-linewidth and frequency-agile photonic integrated extended-DBR lasers}
	%\title{Piezo-tunable photonic integrated circuit based distributed bragg reflector laser with Hertz linewidth}
	\author{Anat Siddharth}
	\affiliation{Laboratory of Photonics and Quantum Measurements, Swiss Federal Institute of Technology Lausanne (EPFL), CH-1015 Lausanne, Switzerland}
	\affiliation{Center of Quantum Science and Engineering, EPFL, CH-1015 Lausanne, Switzerland}
	
	\author{Alaina Attanasio}
	\affiliation{OxideMEMS Lab, Purdue University, 47907 West Lafayette, IN, USA}
	
	\author{Grigory Lihachev}
	\affiliation{Laboratory of Photonics and Quantum Measurements, Swiss Federal Institute of Technology Lausanne (EPFL), CH-1015 Lausanne, Switzerland}
	\affiliation{Center of Quantum Science and Engineering, EPFL, CH-1015 Lausanne, Switzerland}
	
	\author{Junyin Zhang}
	\affiliation{Laboratory of Photonics and Quantum Measurements, Swiss Federal Institute of Technology Lausanne (EPFL), CH-1015 Lausanne, Switzerland}
	\affiliation{Center of Quantum Science and Engineering, EPFL, CH-1015 Lausanne, Switzerland}
	
	\author{Zheru Qiu}
	\affiliation{Laboratory of Photonics and Quantum Measurements, Swiss Federal Institute of Technology Lausanne (EPFL), CH-1015 Lausanne, Switzerland}
	\affiliation{Center of Quantum Science and Engineering, EPFL, CH-1015 Lausanne, Switzerland}
	
	\author{Scott Kenning}
	\affiliation{OxideMEMS Lab, Purdue University, 47907 West Lafayette, IN, USA}
	
	\author{Rui Ning Wang}
	\affiliation{Laboratory of Photonics and Quantum Measurements, Swiss Federal Institute of Technology Lausanne (EPFL), CH-1015 Lausanne, Switzerland}
	\affiliation{Center of Quantum Science and Engineering, EPFL, CH-1015 Lausanne, Switzerland}
	
	\author{Sunil A. Bhave}
	%\email[]{sunil.bhave@epfl.ch}
	\affiliation{OxideMEMS Lab, Purdue University, 47907 West Lafayette, IN, USA}
	
	\author{Johann Riemensberger}
	\email[]{johann.riemensberger@epfl.ch}
	\affiliation{Laboratory of Photonics and Quantum Measurements, Swiss Federal Institute of Technology Lausanne (EPFL), CH-1015 Lausanne, Switzerland}
	\affiliation{Center of Quantum Science and Engineering, EPFL, CH-1015 Lausanne, Switzerland}
	
	\author{Tobias J. Kippenberg}
	\email[]{tobias.kippenberg@epfl.ch}
	\affiliation{Laboratory of Photonics and Quantum Measurements, Swiss Federal Institute of Technology Lausanne (EPFL), CH-1015 Lausanne, Switzerland}
	\affiliation{Center of Quantum Science and Engineering, EPFL, CH-1015 Lausanne, Switzerland}
	\medskip
	\maketitle
	
	%%%%%%%%%%%%%%%%%%%%%%%%%%%%%%%%%%%%%%%%%%%%%%%%%%%%%%%%%%%%%%%%%%%%%%%%%%%%%%%%%%%%%%%%%%%%%%%%%%%%%%%%%%%%%%%
	%%%%%%%%%%%%%%%%%%%%%%%%%%%%%%%%%%%%%%%%%%%%%%%%% Abstract %%%%%%%%%%%%%%%%%%%%%%%%%%%%%%%%%%%%%%%%%%%%%%%%%%%%
	%%%%%%%%%%%%%%%%%%%%%%%%%%%%%%%%%%%%%%%%%%%%%%%%%%%%%%%%%%%%%%%%%%%%%%%%%%%%%%%%%%%%%%%%%%%%%%%%%%%%%%%%%%%%%%%
	
	\noindent\textbf{Recent advances in the development of ultra-low loss silicon nitride (\SiN)-based photonic integrated circuits have allowed integrated lasers to achieve a coherence exceeding those of fiber lasers \cite{li2021reaching}, and enabled unprecedentedly fast (Megahertz bandwidth) tuning using monolithically integrated piezoelectrical actuators \cite{lihachev2022low}. 
	%This has been achieved by self-injection locking of distributed feedback (DFB) laser diodes to silicon nitride (\SiN) microresonators \cite{lihachev2022low, Jin2021}. 
	While this marks the first time that fiber laser coherence is achieved using photonic integrated circuits, in conjunction with frequency agility that exceeds those of legacy bulk lasers, the approach is presently compounded by the high cost of manufacturing DFB, as required for self-injection locking \cite{lihachev2022low,Jin2021}, as well as the precise control over the laser current and temperature to sustain a low noise locked operation.
	%This approach is compounded by the high cost of manufacturing DFB, as well as by the precise control over the laser current and temperature to sustain a low noise and locked operation.
	%The latter can be addressed by using integrated lasers based on reflective semiconductor optical amplifiers (RSOA) in conjunction with Vernier ring reflectors \cite{Tran:2020,lihachev2023frequency} or extended Bragg gratings (E-DBR) \cite{xiang_ultra-narrow_2019}. %, yet these approaches have not allowed for achieving similar performance in coherence or frequency agility. 
	%While this scheme has been utilized for photonic integrated lasers, so far, 
	Reflective semiconductor optical amplifiers (RSOA) provide a cost-effective alternative solution, but have not yet achieved similar performance in coherence or frequency agility, as required for frequency modulated continuous wave (FMCW) LiDAR \cite{Behroozpour:17}, laser locking in frequency metrology \cite{jiang2011making} or wavelength modulation spectroscopy for gas sensing \cite{cassidy1982atmospheric}.
	% no cost-effective RSOA-based integrated lasers exist that are low noise and simultaneously feature fast, mode-hop-free, and linear frequency tuning as required for frequency modulated continuous wave (FMCW) LiDAR \cite{Behroozpour:17}, laser locking in frequency metrology \cite{jiang2011making} or wavelength modulation spectroscopy for gas sensing \cite{cassidy1982atmospheric} have been demonstrated to date.
	Here, we overcome this challenge and demonstrate an RSOA-based and frequency agile integrated laser that can be tuned with high speed, good linearity, high optical output power, and turn-key operability while maintaining a small footprint.
	This is achieved using a tunable extended distributed Bragg reflector (E-DBR) in an ultra-low loss 200~nm thin \SiN platform with monolithically integrated piezoelectric actuators \cite{liu2020monolithic,tian2020hybrid}.
	%, that allows linear, low power, and hysteresis free tuning.% to demonstrate an RSOA-based and frequency agile integrated laser that can be tuned with high speed, good linearity, high optical output power, and turn-key operability.
	We co-integrate the DBR with a compact ultra-low loss spiral resonator to further reduce the intrinsic optical linewidth of the laser to the Hertz-level - on par with the noise of a fiber laser - via self-injection locking.
The photonic integrated E-DBR lasers operate at 1550~nm, feature up to 25~mW fiber-coupled output power in the free-running and up to 10.5 mW output power in the self-injection locked state. 
The intrinsic linewidth is 2.5~kHz in the free running state and as low as 4~Hz in the self-injection locked state.
In addition, we demonstrate the suitability for FMCW LiDAR by showing laser frequency tuning over 1.0~GHz at up to 1~MHz triangular chirp rate with a nonlinearity of less than 0.4\% without linearization by modulating Bragg grating using monolithically integrated Aluminum Nitride (AlN) piezoactuators.    
	}
	
	%%%%%%%%%%%%%%%%%%%%%%%%%%%%%%%%%%%%%%%%%%%%%%%%%%%%%%%%%%%%%%%%%%%%%%%%%%%%%%%%%%%%%%%%%%%%%%%%%%%%%%%%%%%%%%%
	%%%%%%%%%%%%%%%%%%%%%%%%%%%%%%%%%%%%%%%%%%%%%%%% Introduction %%%%%%%%%%%%%%%%%%%%%%%%%%%%%%%%%%%%%%%%%%%%%%%%%
	%%%%%%%%%%%%%%%%%%%%%%%%%%%%%%%%%%%%%%%%%%%%%%%%%%%%%%%%%%%%%%%%%%%%%%%%%%%%%%%%%%%%%%%%%%%%%%%%%%%%%%%%%%%%%%%
	
	\begin{figure*}[htbp!]
		\centering
		\includegraphics[width=1\linewidth]{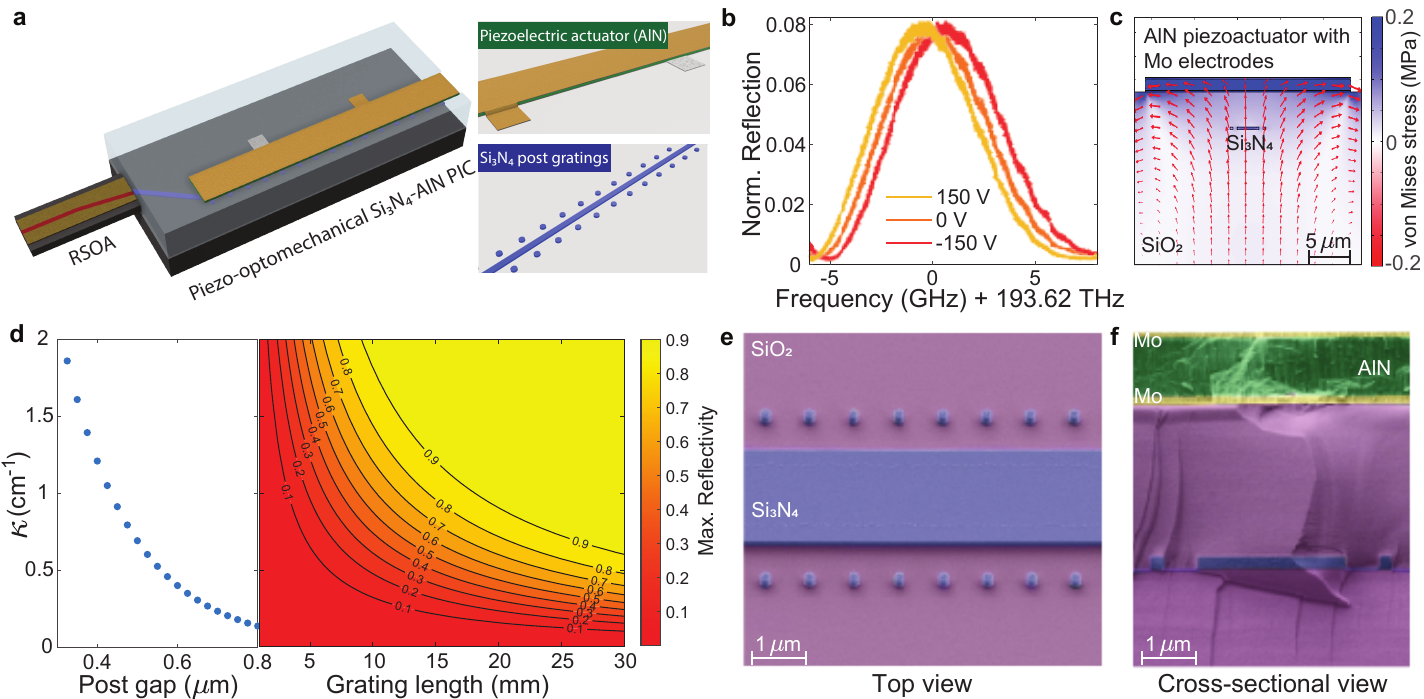}
		\caption{\textbf{Concept and device.}
			(a) Schematic illustration of the hybrid integrated extended-DBR (E-DBR) laser. An RSOA is butt-coupled to the piezo-optomechanical chip comprising Bragg gratings of \SiN and AlN as piezoelectric material with Mo as top and bottom metal electrodes.
			(b) Voltage tuning of the center frequency of the $\kappa$=0.6 cm$^{-1}$ grating.
			(c) Vertical stress distribution in the \SiN-AlN photonic chip on applying 1~V across the Mo electrodes.
			(d) Simulated coupling strength of the uniform second-order Bragg gratings. The plot on the right shows the variation of peak reflectivity of such gratings with changes in coupling strength and the length of the grating. 
			(e) False-colored Scanning Electron Microscope (SEM) images of the top and (f) cross-section view of the MEMS-Photonic chip.}
		\label{fig1}
	\end{figure*}
	
	\begin{figure*}[htbp!]
		\centering
		\includegraphics[width=1\linewidth]{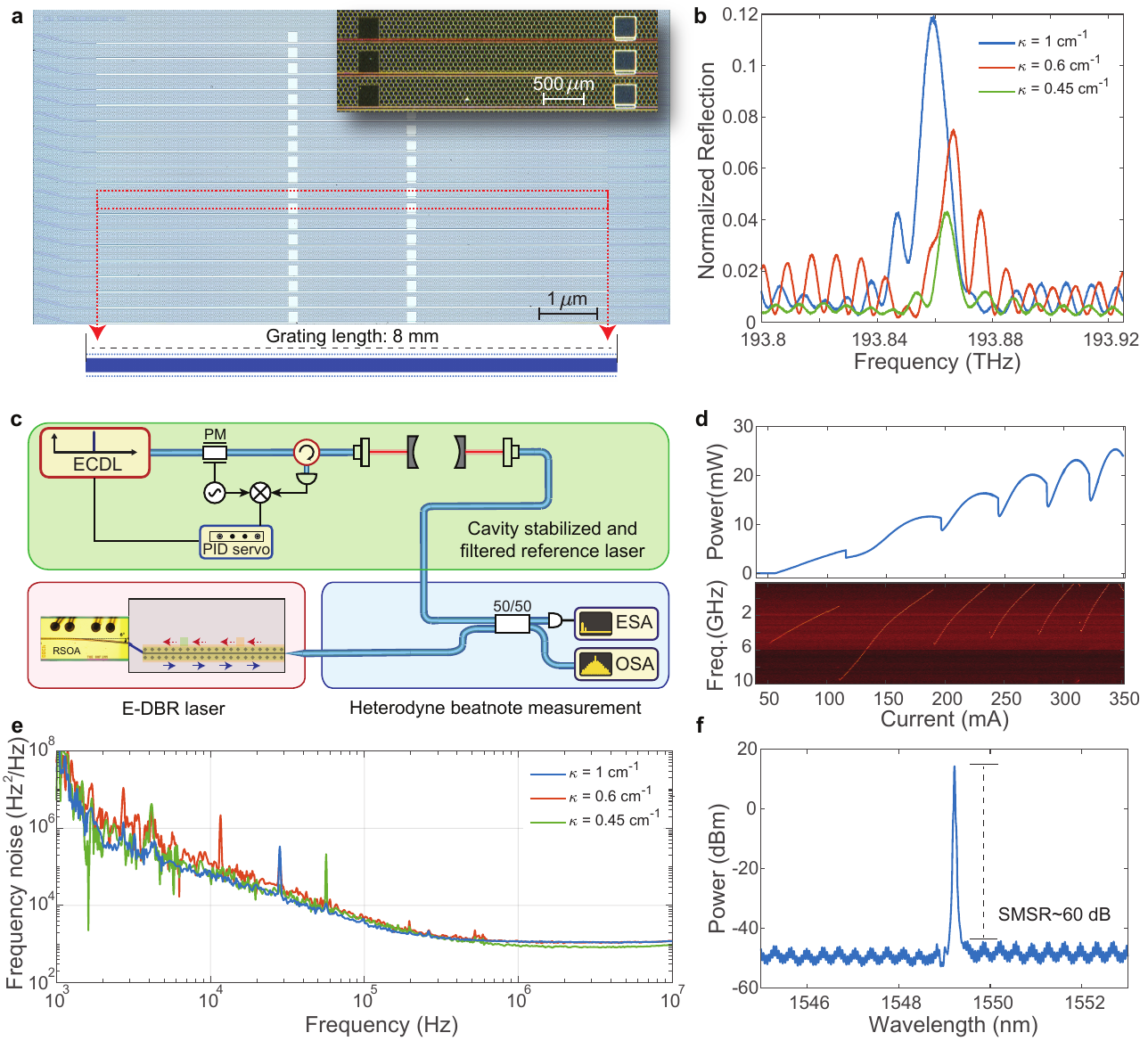}
		\caption{\textbf{Performance of laser with external distributed Bragg reflector (E-DBR).}
			(a) Photograph of the photonic chip (\texttt{D116\_02\_F9\_C9}) comprising of the 8~mm long E-DBR with monolithically integrated AlN as piezoactuator. The inset shows a zoomed-in image of the E-DBR with electrodes of the piezoactuator fabricated via the 'Pull-back' process.
			(b) Measured reflection of 2$^{\mathrm{nd}}$-order Bragg gratings with different post distances and simulated grating strength.
			(c) Schematic measurement setup for laser frequency noise measurement by heterodyne beat spectroscopy with an FP-cavity stabilized and filtered ECDL. 
			(d) Top: Laser power as a function of RSOA injection current. Bottom: Heterodyne frequency tuning of laser with injection current modulation.
			(e) Frequency noise of E-DBR lasers with different reflection gratings.
			(f) Optical spectrum of E-DBR laser featuring more than 60 dB side mode suppression ratio.}
		\label{fig2}
	\end{figure*}
	
	\begin{figure*}[htbp!]
		\centering
		\includegraphics[width=1\linewidth]{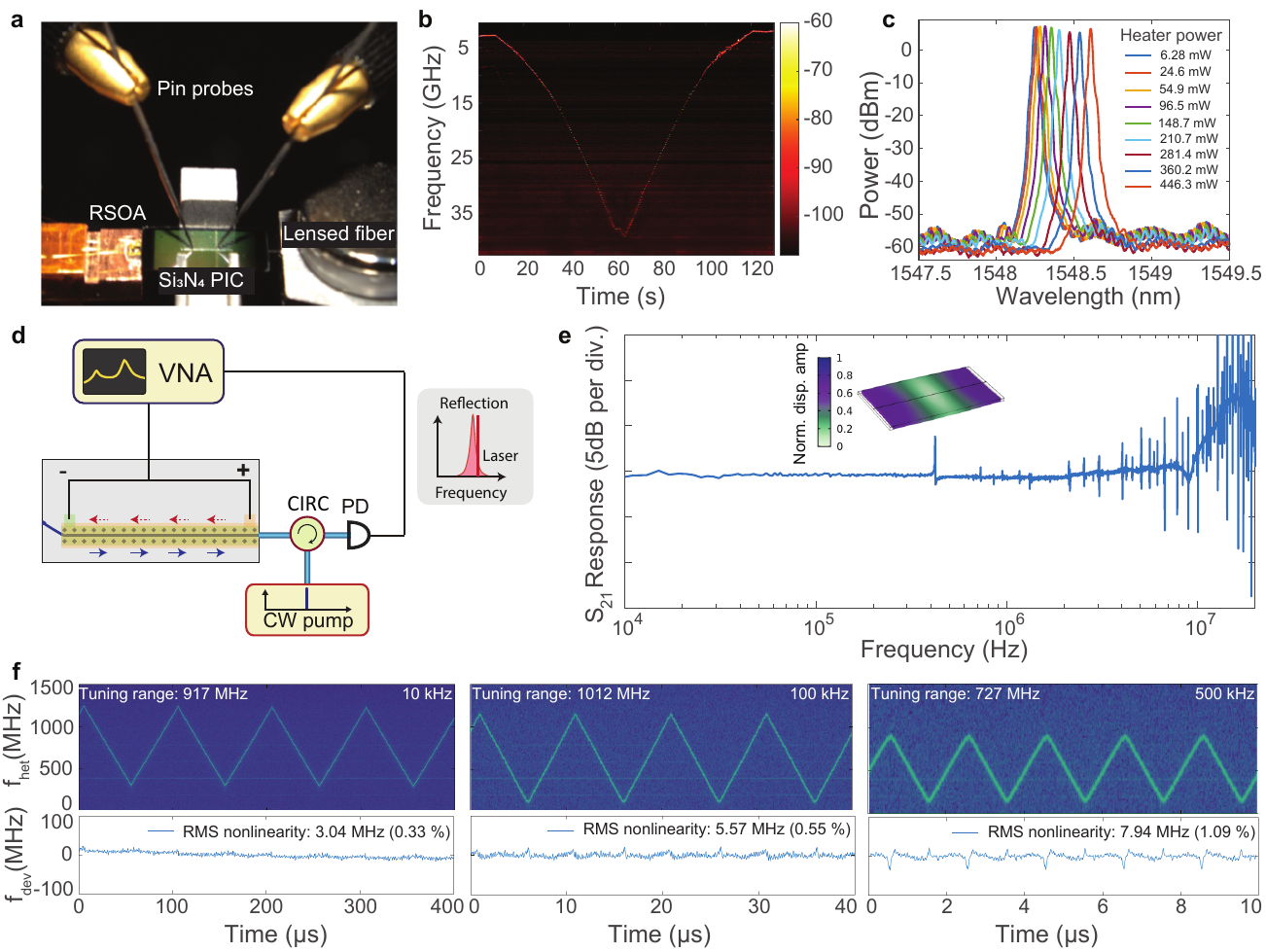}
		\caption{\textbf{Frequency tuning of the E-DBR laser.}
			(a) Photo of optical setup used for E-DBR laser tuning.
			(b) Heterodyne beat spectrogram of laser tuning using an optical microheater showing mode-hop free tuning range over 40~GHz. 
			(c) Optical spectrum of laser for different microheater powers.
			(d) Measurement setup for response measurement of piezoelectric tuning. The external laser (ECDL) is tuned to the side of the grating reflection peak and the reflected light is isolated using an optical circulator and the tuning response is recorded using a vector network analyzer (VNA).
			(e) Piezoelectric tuning response of the DBR grating. The displacement field of the first FBAR mode at 400~kHz is plotted in the inset.
			(f) Heterodyne beat spectrograms of laser tuning applying 250~V$_\mathrm{pp}$ (peak-to-peak) triangular waveforms at 10 kHz (left), 100~kHz (middle), and 500~kHz (right) across the AlN actuator.}
		\label{fig3}
	\end{figure*}
	
	\begin{figure*}[htbp!]
		\centering
		\includegraphics[width=1\linewidth]{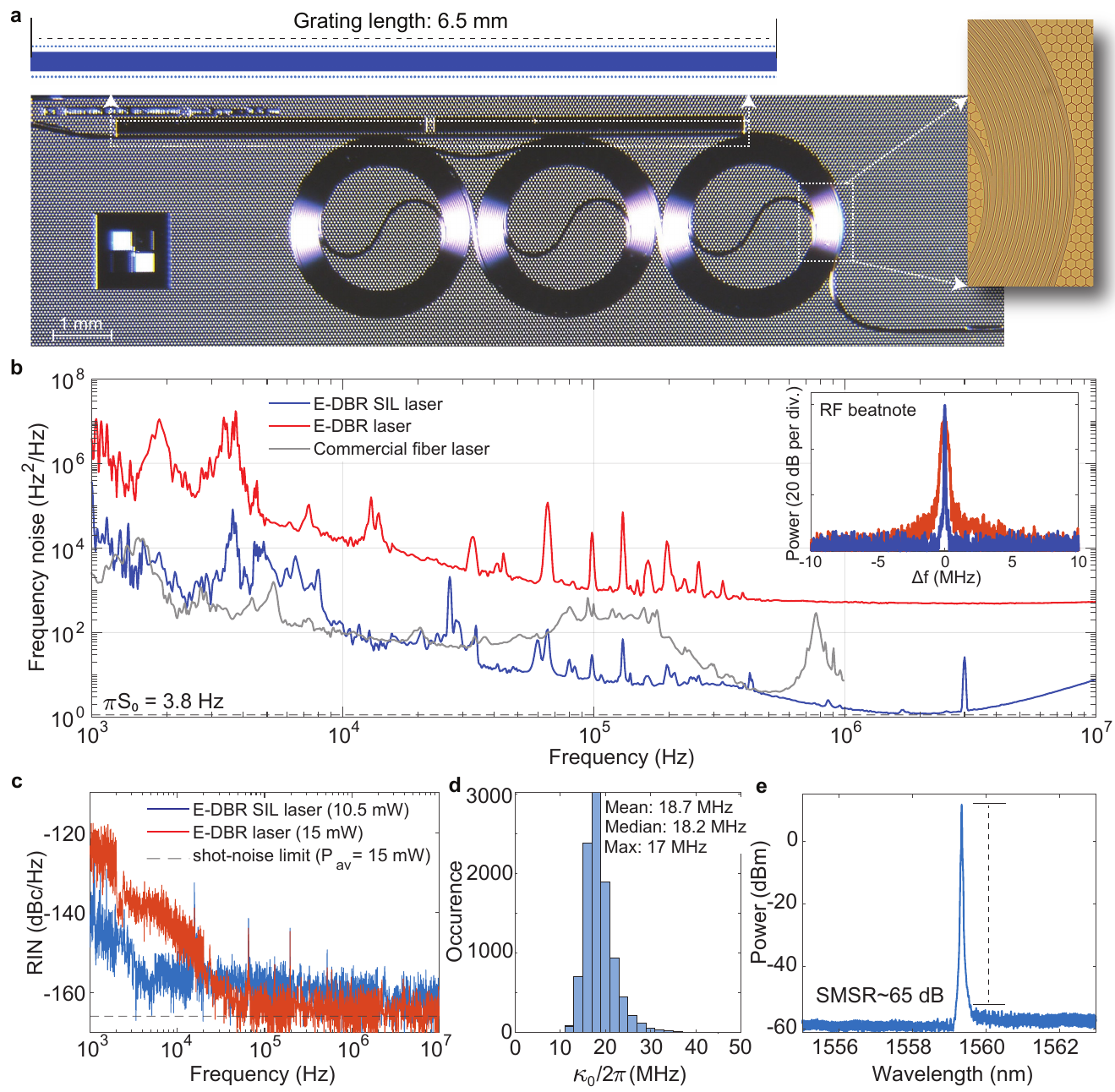}
		\caption{\textbf{Noise characterization of the E-DBR laser self-injection locked to a high-Q 800~MHz spiral cavity.}
			(a) Photo of the photonic chip (\texttt{D128\_01\_F7\_C11\_1\_01}) comprising of a 6.5~mm long E-DBR and a compact and ultra-low loss spiral resonator with 800~MHz free-spectral range(FSR). The inset shows the pulley coupler for coupling light to the high-Q spiral resonator.
			(b) Optical cavity loss rate histogram of the spiral resonator.
			(c) Single sideband power spectral density of frequency noise of E-DBR laser in the unlocked and the self-injection-locked regimes. Inset: Heterodyne beat note spectrum (RBW: 100 kHz).
			(d) Relative intensity noise (RIN) of E-DBR laser in the free-running (blue) and the self-injection locked regime (red).
			(e) Optical spectrum of E-DBR laser in the self-injection locked regime achieving more than 65 dB SMSR (RBW: 0.02 nm).}
		\label{fig4}
	\end{figure*}
	
	\section{Introduction}
	The use of narrow linewidth lasers has become widespread due to their multiple applications in fields such as metrology \cite{guo2022chip}, optical sensing \cite{rogers1999distributed}, microwave photonics \cite{marpaung2019integrated}, addressing atomic and molecular quantum systems such as atomic clocks \cite{jiang2011making}, quantum computers \cite{pogorelov2021compact}, and coherent communications for metro- and long-haul interconnects. 
	The ability to tune and control laser frequency is a prerequisite for laser locking in numerous quantum physics experiments, fast wavelength switching in telecommunications \cite{guan2018widely}, and frequency-modulated continuous-wave LiDAR \cite{Behroozpour:17,riemensberger2020massively}.
	While silicon-based integrated lasers have become more compact and commercially used in data center transceivers, their phase noise performance so far remains inferior to continuous-wave fiber lasers.
	E-DBR lasers with feedback circuits implemented in photonic integrated circuits demonstrated significant progress in recent years \cite{Xiang2021, Belt:14, Huang:19,tran2019tutorial}.
	The low propagation loss and low defect densities of \SiN integrated photonic circuits \cite{liu2021high} have enabled novel functionalities such as traveling wave parametric amplifiers \cite{Riemensberger2022}, Erbium-doped amplifiers \cite{liu2022photonic} or Erbium lasers \cite{liu2023fully} along with meter-long spiral resonators that have led to a new class of integrated lasers that can, for the first time, surpass the coherence of fiber lasers in terms of phase noise \cite{Jin2021, Li:21}.
	Low loss \SiN PICs have also been integrated monolithically with piezoelectric actuators - using both Aluminum Nitride (AlN) and Lead Zirconate Titanate (PZT) \cite{liu2020monolithic, dong2018port, lihachev2022low, lihachev2023frequency}, allowing for fast (> 1 MHz), low-power, low hysteresis and linear tuning, resulting in energy-efficient phase shifters \cite{everhardt2022ultra}, frequency-agile low-noise lasers \cite{lihachev2022low},  fast-tunable soliton frequency combs \cite{liu2020monolithic} or non-reciprocal devices \cite{tian2021magnetic}.
	
	Here, we demonstrate a high-power ultra-low noise laser with rapid non-thermal and highly linear tuning capability based on the monolithic integration of AlN piezoelectric thin films on top of an \SiN waveguide Bragg grating. 
	We achieve a GHz-level tuning range at up to 1~MHz triangular chirp rate with an RMS nonlinearity of less than 0.4$\%$, limited by the available electronics and without the need for signal predistortion \cite{Feneyrou:17} at a fiber-coupled output power of 25~mW.
	We also demonstrate that the E-DBR laser can be self-injection locked to an ultra-low loss spiral resonator with a small free-spectral range of 800~MHz on the same photonic chip with a footprint of only 20~mm$^2$ and achieve an intrinsic linewidth of less than 4~Hz and a RIN of -160~dBc/Hz at 100~kHz offset and output power of more than 10~mW.
	
	\section{E-DBR Design and Characterization}
	
	As illustrated in Figure \ref{fig1} the E-DBR laser is comprised of a reflective semiconductor optical amplifier (RSOA) and a \SiN PIC with a monolithically integrated AlN-based piezoelectric actuator.
	A Bragg grating with low coupling strength $\kappa$ is needed to attain a narrowband reflection for E-DBR operation \cite{xiang_ultra-narrow_2019}. 
	
	We fabricate 8~mm long E-DBR gratings in a 200~nm thick subtractive \SiN platform with a 2.5~$\mu m$ wide waveguide. 
	The grating is formed by circular \SiN posts placed on either side of the waveguide \cite{spencer_low_2015}.
	We designed the \SiN Bragg gratings as second-order gratings because of the resolution limits of our DUV lithography system. 
	We set the duty cycle to 35\% presenting a strong second-order resonance with grating period $\Lambda$=997.1~nm and post diameter 350~nm as shown in the SEM images in Figure \ref{fig1}(e,f). 
	We analyzed the Fourier components of the circular grating posts in the second order period to calculate the coupling strength of the forward and backward components in the grating as a function of the distance between the waveguide and post sidewalls (cf. Supplementary Information). The coupling strength of the designed \SiN gratings with its peak reflectivity for different grating lengths is shown in Figure \ref{fig1}d.  
	The size and spacing of the grating posts are kept constant as no apodization is required in weakly reflecting E-DBR gratings to suppress side maxima to the level of the chip facet reflections. 
	To make the PIC grating voltage-tunable, we monolithically integrate piezo-electrical actuators based on a 1~$\mu$m AlN thin film \cite{dong2018port,liu2020monolithic} with top and bottom electrodes formed from 100~nm Molybdenum (Mo) films. 
	The voltage tuning of the reflection from the Bragg gratings is shown in Figure \ref{fig1}b and the vertical stress distribution on applying 1 Volt to the AlN piezoactuator is shown in Figure \ref{fig1}c.
	The AlN piezoactuators with Molybdenum (Mo) top and bottom electrodes are placed on top of a 3.5 $\mu$m SiO$_2$ cladding to minimize the metal-induced optical loss.      
	
	Figure \ref{fig2}b depicts the optical measurement of the grating reflection ranging between 4\% and 12\%, which we measured using a lensed fiber and external cavity diode laser (ECDL).  
	The input coupling between the chip and the RSOA is facilitated by a 4.5~$\mu$m wide and 300~$\mu$m long linear horn taper. 
	The taper angle of 12.4$^{\circ}$ is chosen to match the emission angle of the RSOA accounting for refraction according to Snell's law. 
	The overlap integral of the laser and the waveguide mode is larger than 80$\%$ (cf. Supplementary Information). 
	The insertion loss from lensed fiber to the 500~nm wide inverse taper is estimated as 1~dB.
	The fabrication process of the \SiN-AlN PIC is described in the Methods section and the fabrication process to fabricate low-capacitance bond pads is explained in the Supplementary Information.

	\section{E-DBR laser operation and noise}
	
	The RSOA is mounted on a 3D piezoelectrical translation stage and butt-coupled to the \SiN PIC as shown in Figure \ref{fig3}a with a nominal maximal output power of 60~mW.
	As the drive current is increased above the threshold of 50~mA, lasing at the grating wavelength of 1549.5~nm occurs.
	The laser power and frequency tuning curves as a function of the drive current are depicted in Figure \ref{fig2}d. 
	As the current is increased to 350~mA, the laser power increases up to 25~mW with a side-mode suppression ratio in excess of 60~dB (cf. Figure \ref{fig2}f), and intermediate mode-hops are observed at drive currents 110~mA, 200~mA, 240~mA, 278~mA, and 320~mA. 
	The laser frequency tunes over the 15 GHz width of the grating, highlighting the function of the RSOA as an intracavity phase shifter. 
	Hence, no additional thermal phase shifter is required on the \SiN chip to operate the laser. 
	We measure the laser frequency noise power spectral density (single-sided PSD, in units of Hz$^2$/Hz) using heterodyne beat spectroscopy with an external-cavity diode laser (ECDL) that is locked to and filtered by a free-space Fabry-P\'erot (FP) cavity with 1.5~GHz FSR and a linewidth of 70~kHz (cf.~Figure \ref{fig2}c).
	For the gratings with strength 1 cm$^{-1}$ and 0.61~cm$^{-1}$, we measured nearly identical frequency noise and intrinsic linewidth of 3.45~kHz and for the grating with strength 0.4~cm$^{-1}$, we measured an intrinsic linewidth of 2.5~kHz.

	\section{Laser frequency tuning with integrated piezoactuator}
	
	Next, we characterize the frequency tuning characteristics of the RSOA-based E-DBR lasers. 
	The first laser grating is configured with Ti/Pt thermal heaters that are fabricated atop SiO$_2$ cladding, with a 2.5 $\mu$m offset from the waveguide.
	This allows for slow but wideband tuning of the E-DBR reflection peak and the laser power.
	We increase the power on the thermal microheater up to 446.3~mW and measure up to 0.5~nm ($>$ 40 GHz) mode-hop-free laser tuning (cf. Figure \ref{fig3}b,c).
	The heater efficiency can be further improved by optimizing the placement of metal heaters with respect to the waveguide. 
	The laser operates in a longitudinal mode across the full tuning range. 
	For fast frequency modulation, we fabricated a \SiN gratings PIC with AlN piezoactuators. 
	We measured the small-signal frequency response of the gratings using an auxiliary laser and a vector network analyzer (VNA).
	Figure \ref{fig3}d shows the setup of the actuation response measurement. 
	In this measurement, the actuation voltage derived from a vector network analyzer (VNA) is applied to the actuator, and a laser is frequency-tuned to sit on the side of the reflection peak, and the phase fluctuation of the grating reflection is measured as amplitude response using a fast photodetector.
	The reflected light is separated via an optical circulator (CIRC) and detected on a photodiode (PD).
	We find that the frequency response is flat up to the 400~kHz frequency, which is the fundamental bulk mode of the 1~cm by 5~mm photonic chip.
	The inset of Figure 3(e) shows finite element simulations of the bulk modes of the photonic chip, that match the observed actuation resonance. 
	At higher frequencies, the density of observable flexural and bulk modes increases and at 17~MHz the first bulk acoustic mode appears. 
	The increasing mode density of the \SiN photonic chip towards higher frequencies limits the flat modulation bandwidth.
	The excitation of spurious mechanical modes of the chip can be avoided using the differential drive of two piezo actuators in opposite phase \cite{lihachev2022low} or by using irregular polygonal shapes for the photonic chips \cite{burak2017acoustic}. 
	Appropriate photonic packaging can also greatly dampen the flexural and bulk modes of the chip.
	These methods can extend the actuation bandwidth of the laser to greater than 10~MHz \cite{siddharth2021actuation} up to the first bulk acoustic resonance of the chip.
	However, for periodic actuation with sinusoidal or triangular waveforms, the actuation of the narrow flexural and bulk modes can also be averted by choice of actuation frequency such that the (harmonic) spectra of the drive frequency do not overlap with such mechanical modes.
	In Figure \ref{fig3}(f), we demonstrate high-speed tuning of the E-DBR laser and show the time-frequency pattern $f_\mathrm{het}$ determined by short-time Fourier transform of the heterodyne beat note of the E-DBR laser and the reference laser with triangular modulation frequencies of 10~kHz, 100~kHz, and 500~kHz. 
	The bottom panel row depicts the deviation of the measured frequency tuning pattern from a perfect triangular frequency modulation with RMS deviations ranging from 0.3\% to 1\% without any predistortion compensation.

	\section{Self-injection locking of E-DBR laser}
	
	In order to further reduce the optical linewidth of the E-DBR laser, we can use the very low waveguide propagation loss of the 200~nm thin \SiN platform and fabricate a high-Q microresonator on the same substrate as the grating and utilize self-injection locking (SIL) to drastically reduce the optical linewidth \cite{lihachev2022low,liang2010whispering,siddharth2022near}. 
	The waveguide width of the spiral resonator is 5~$\mu m$ to decrease the field overlap with the rough sidewalls and mitigate scattering loss \cite{bauters2011ultra}.
	The spiral resonator is comprised of three archimedean spirals connected with curvature and curvature-derivative continuous splines \cite{chen2012general} and coupled to the bus waveguide using pulley couplers as shown in Figure \ref{fig4}a. 
	We eliminate higher order modes in the resonators by inserting 1.5 $\mu m$ wide tapering sections in the central S-bend of each spiral.
	The median intrinsic cavity linewidth $\kappa_0/2\pi$ of the \SiN cavity is 18.2~MHz, which corresponds to a linear loss of 2.5~dB/m. 
	Since the laser linewidth narrowing factor, i.e. the ratio of the free-running laser linewidth to the linewidth of the injection-locked laser is proportional to the square of the loaded quality factor of the resonator mode to which the laser is self-injection-locked\cite{galiev2018spectrum}, the high loaded Q of the \SiN spiral microresonator (cf. Figure \ref{fig4}d) can significantly reduce the optical linewidth and improve the side mode suppression ratio (SMSR), which increases from 60~dB to 65~dB (cf. Figure \ref{fig4}e). 
	We operate the spiral microresonator in the undercoupled ($\kappa_{ex}$=2~MHz) regime to minimize the power loss of the E-DBR laser system in the locked state and hence achieve an output power in the fiber of 10.5~mW compared to 15~mW in the unlocked state. The loaded quality factor of the resonance used for self-injection locking is 10 $\times$ 10$^6$.
	Locking is achieved by tuning the RSOA drive current to tune the laser within the 10~GHz FWHM reflection bandwidth of the grating. 
	The dense mode spacing of the long spiral resonator ensures frequency overlap between the grating and the resonator. 
	We measure the frequency noise power spectral density (PSD) of the self-injection locked E-DBR laser by employing the same technique as described previously (cf. Figure \ref{fig2}c).
	Here, the gratings are 6.5~mm long and have coupling strength of 0.6~cm$^{-1}$ and the frequency noise PSD before (red trace) and after (blue trace) self-injection locking to the spiral cavity is shown in Figure \ref{fig4}b.
	As can be seen, the laser frequency noise is decreased by more than 20~dB over the whole frequency range, resulting in a white noise floor below 1.5 Hz$^2$/Hz and an intrinsic linewidth of about 3.8~Hz. 
	The inset shows the narrowing of the heterodyne beatnote spectrum upon self-injection locking. The grey trace shows the frequency noise PSD of a commercial fiber laser and thus verifies that the self-injection locked E-DBR laser reaches fiber laser coherence (cf. Supplementary Information).
	The relative intensity noise (RIN) spectrum of the laser in the locked and unlocked state is presented in Figure \ref{fig4}(c).
	Fabricating both the grating and the optical microresonator on the same chip improves the relative stability of both elements and the stability of the laser overall.
	We measure laser intensity noise close to the shot noise limited regime for frequencies larger than 100~kHz, which reaches a RIN of -160~dBc/Hz regardless of operation in the self-injection locking or free-running regime of the E-DBR.
	
	\section*{Conclusion}
	
	Our E-DBR laser demonstrates the ability to manufacture frequency-agile lasers without relying on expensive high-power DFB laser diodes by transferring the frequency selective element of an ultra-low noise SIL laser from the III-V chip to the \SiN PIC with increased fabrication tolerance due to the low index and higher order E-DBR grating. 
	We achieve high-power operation with output powers up to 25~mW and GHz-level tuning range and up to 500~kHz rate. 
	By self-injection locking of the E-DBR laser to an ultra-low loss spiral resonator, we achieve an intrinsic linewidth of 3.8~Hz at an unprecedented output power of 10.5~mW for hybrid lasers with a total footprint of the combined \SiN gratings and spiral resonator of merely 20 mm$^2$, which is considerably smaller than previous demonstrations using \SiN with thickness below 100~nm \cite{Jin2021, Li:21}.
	Placing the grating and the spiral resonator on the same substrate and in close proximity also greatly improve the temperature stability of the combined system, where SIL lasers have hitherto required temperature control in the millikelvin range for stable operation \cite{lihachev2022low,li2021reaching}. 
	In summary, our laser system is an ideal solution for applications in frequency metrology, fiber sensing, or spaceborne LiDAR, where high power, exceptional coherence, and fast frequency modulation capabilities are required. 
	
	\section*{Methods}
	
	\subsection{Fabrication of the \SiN-AlN photonic chips}
	The \SiN photonic chips are fabricated with a custom subtractive waveguide manufacturing process based on 248~nm DUV stepper lithography with fluoride-based dry reactive ion etching chemistry. 
	\SiN thin films of 200~nm are procured externally and waveguide cores are etched using a fluoride-based dry reactive ion etching chemistry.
	After etching we grow a 10~nm thin layer of \SiN to conformally cover the waveguides using low-pressure chemical vapor deposition to reduce sidewall roughness \cite{torres2022ultralow} and absorption by surface located defect states induced in the etching \cite{puckett2021422}. 
	A high-temperature anneal (11~h, 1200$^{\circ}$) is conducted before a 3~$\mu$m top oxide cladding is deposited. 
	The AlN actuators are fabricated with the same process used in Ref.\cite{tian2020hybrid}. The electrodes are fabricated using a 'Pull-back' process described in the Supplementary Information.
	After chip release, a low temperature (30~min,300°C)) annealing cycle is conducted to remove long-lived defect states induced by deep-UV exposure.
	
	\subsection{Relative intensity noise measurement (RIN)}
	The laser relative intensity noise was measured using direct photodetection of the laser output at 1550~nm by an impedance-matched photodetector (PD, XPDV2120RA, responsivity $\gamma_{PD}=$0.63~A/W).  
	%During the measurement, the E-DBR laser output was tapped out for the inspection of the optical spectrum and temporal trace.
	In the measurement, the input power to the photodetector from the E-DBR and the E-DBR SIL laser was recorded as 15~mW and 10.5~mW, respectively.
	The output of the photodetector was amplified by an RF amplifier (MITEQ AM-1676) to lift the detected noise above the displayed average noise level (DANL) of the electrical spectrum analyzer (ESA, R\&S, FSW44). 
	The gain response $G(f)$ of the RF amplifier was characterized using a vector network analyzer with $-45$~dBm output power to avoid gain saturation.
	The total noise power spectrum density (PSD) at the PD output is given by
	\begin{equation}
		\Delta P(f)=\frac{P_\mathrm{N}(f)}{G(f) \cdot \mathrm{RBW}},
		\label{eqn:RIN power}
	\end{equation}
	where $P(f)$ is the noise power measured by ESA when the amplified signal is sent to the PD.
	The shot noise $S_\mathrm{shot}(f) = 2h\nu\mathrm{\bar{P}}$ is calculated based on the average output power.
	The laser RIN is then given by
	\begin{equation}
		\mathrm{RIN}=10 \log _{10} \frac{\Delta P(f)}{I_\mathrm{av}^{2} R_\mathrm{load}}.
		\label{eqn:RIN}
	\end{equation}
	%For the measurement presented in this work (Fig.\ref{Fig:RIN noise}), we focused on the frequency range within 400~MHz which is important for most of the applications such as coherent sensing and optical communications.
	%Note that in this measurement the low RIN noise at high frequencies might be comparable with the shot noise, and it can lead to inaccuracy over a certain range which should be abandoned in the figure presentation.
	%To overcome this limitation, an increased photocurrent or optical amplification is needed.
	% In the latter case, the excess RIN from an optical amplifier needs to be characterized or is usually neglected. 
	RIN at higher frequencies can be measured using an RF amplifier with a broader bandwidth, or without using an RF amplifier when the photocurrent is sufficiently high to overcome the DANL and the shot noise.
	
	\subsection{Fabry-P\'erot filtering cavity}
	The cavity is assembled using the 10CV00SR.70F High-Performance Concave SuperMirrors$^{TM}$ from Newport. The mirrors are glued to the ZERODUR ultra-low expansion (ULE) glass spacer with the coating at the inner side such that they would form a stable symmetric optical cavity. The ULE glass is used to ensure the thermal stability of the cavity length because the cavity length resolution could be very high. A piezo (Piezosystem Jena HH1-2515-07) is also integrated into this assembly in order to tune the cavity resonance frequency by changing the cavity length, and will eventually limit the cavity thermal stability. 
	The concave mirrors with the piezo form a very planar cavity with Finesse $\mathcal{F}$, FSR 2.14 GHz, and linewidth $\Delta \nu_{1/2} = \mathrm{FSR}/\mathcal{F} = 71.4$ kHz. 
	The goal is to have a narrow linewidth, high transmission cavity at 1550~nm to lock the laser and filter the high-frequency noise.
	
	\begin{footnotesize}
		
		%\noindent \textbf{Supplementary Material}: 

		\noindent \textbf{Author Contributions}:
		A.S., J.R. simulated and designed the devices.
		R.N.W., Z.Q fabricated the device. 
		A.A. fabricated integrated actuators.
		A.S. and G.L. characterized the devices. 
		A.S. did the experiments and analyzed the data with the help of J.R., J.Z. and G.L. A.S., J.R., and G.L. wrote the manuscript with input from all authors. 
		T.J.K., J.R., and S.B. supervised the project.
		
		\noindent \textbf{Funding Information and Disclaimer}: This publication was supported by Contract W911NF2120248 (NINJA LASER) from the Defense Advanced Research Projects Agency (DARPA), Microsystems Technology Office (MTO), as well as the Swiss National Science Foundation (SNSF) through grant number 211728 (BRIDGE).
		A.S. acknowledges support from the European Space Technology Centre with ESA Contract No. 4000135357/21/NL/GLC/my and J.R. acknowledges support from the Swiss National Science Foundation under grant no. 201923 (Ambizione).
		
		\noindent \textbf{Acknowledgments}:
		The authors thank Guanhao Huang for assembling the Fabry-P\'erot filtering cavity.
		The chip samples were fabricated in the EPFL center of MicroNanoTechnology (CMi), and in the Birck Nanotechnology Center at Purdue University.
		
		\noindent \textbf{Data Availability Statement}: The code and data used to produce the plots within this work will be released on the repository \texttt{Zenodo} upon publication of this preprint.
		
		\noindent\textbf{Correspondence and requests for materials} should be addressed to T.J.K.
	\end{footnotesize}
	\bibliography{citations}
	
\end{document}

%\includepdf[landscape=false]{Science_SM_Cover.pdf}
\title{Supplementary Information for: Hertz-linewidth and frequency-agile photonic integrated extended-DBR lasers}

\author{Anat Siddharth$^{1,2}$,
				Alaina Anastasio$^{3}$, 
			    Grigory Lihachev$^{1,2}$, 
			    Junyin Zhang$^{1,2}$, 
			    Zheru Qiu$^{1,2}$, 
			    Scott Kenning$^{3}$,
			    Rui Ning Wang$^{1,2}$, 
			    Sunil Bhave$^{3}$,
			    Johann Riemensberger$^{1,2,\ddag}$,
				and Tobias J. Kippenberg$^{1,2,\ddag}$}
\affiliation{
$^1$Laboratory of Photonics and Quantum Measurements, Swiss Federal Institute of Technology Lausanne (EPFL), CH-1015 Lausanne, Switzerland\\
$^2$Center for Quantum Science and Engineering, EPFL, CH-1015 Lausanne, Switzerland\\
$^3$OxideMEMS Lab, Purdue University, 47907 West Lafayette, IN, USA
}

%%%%%% RESET EQUATION NUMBERS ETC. %%%%%%%%
\setcounter{equation}{0}
\setcounter{figure}{0}
\setcounter{table}{0}

\setcounter{subsection}{0}
\setcounter{section}{0}
\setcounter{secnumdepth}{3}

%
%\begin{abstract}
%\hl{JR: Would remove, table of contents suffices! Supplementary Information accompanying the manuscript containing analysis of grating strength, laser performance comparison with state-of-the-art integrated lasers and fiber lasers, Pull-back fabrication process of electrodes, bending and metal-induced propagation loss analysis, Optical Frequency Domain Reflectometry (OFDR) measurement of a long spiral waveguide in the same 200~nm thin \ce{Si3N4} platform to determine propagation loss and edge-coupling simulation to design optimized tapers for efficient coupling of light from RSOA.} 
%\end{abstract}

\maketitle
{\hypersetup{linkcolor=blue}\tableofcontents}
\newpage

%%%%%%%%%%%%%%%%%%%%%%%%%%%%%%%%%%%%%%%%%%%%%%%%%%%%%%%%%%%%%%%
\section{Analysis of the grating strength}
\begin{figure*}[htb]
	\centering
	\includegraphics[width=1\textwidth]{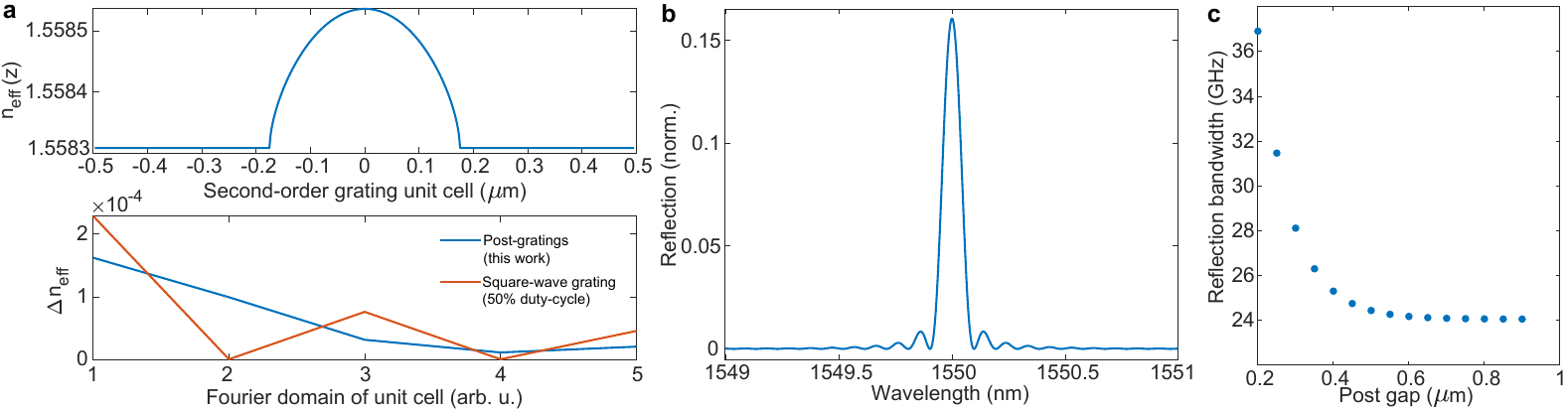}
	\caption{ 
		\footnotesize
		\textbf{Fourier analysis of the coupling strength.} (a) Distribution of the effective refractive index in one period of the Bragg grating (top) and $g_\mathrm{d}=350$ nm with and the change in effective refractive index with respect to the Fourier domain of the grating unit cell (bottom). (b) Simulation of the reflection spectra with $\kappa$=1 cm$^{-1}$. (c) Simulated reflection bandwidth of gratings with different coupling strength, varied by sweeping the gap distance between the posts and the waveguide.}
	\label{Fig:fft}
\end{figure*}

\noindent Bragg gratings can be conceptualized as one-dimensional diffraction gratings which diffract light from the forward-travelling mode into the backward-travelling mode. The reflections from succeeding periods of the grating must constructively interfere in order for light to be efficiently diffracted in the opposite direction, hence the optical length between succeeding reflection interfaces must be an integer multiple number of the vacuum wavelength in order to interfere constructively. The Bragg condition for such a m$^{th}$-order grating is
\begin{equation}
\Lambda = \frac{m\lambda_\mathrm{B}}{2n_\mathrm{eff}},
\end{equation}
where $\Lambda$ is the Bragg period and $\lambda_\mathrm{B}$ is the free-space Bragg wavelength.
\begin{figure*}[htb]
	\centering
	\includegraphics[width=0.75\textwidth]{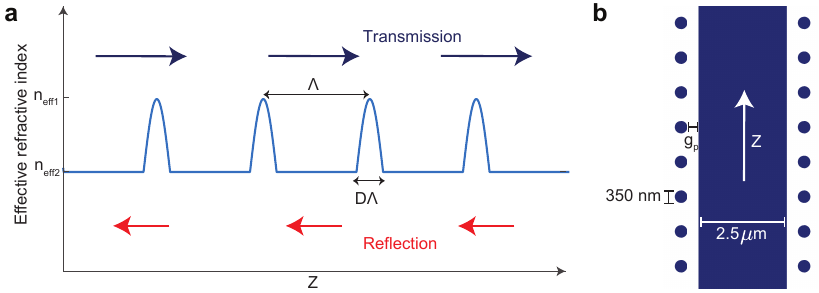}
	\caption{ 
		\footnotesize
		\textbf{Bragg gratings.} (a) The graph illustrates longitudinal effective index profile of a uniform grating with $\Lambda$ as the grating period and $D$ as the duty-cycle. (b) Schematic of the \ce{Si3N4} post-grating design.}

	\label{Fig:GDS layout}
\end{figure*}
\noindent If we assume, the periodic variation of refractive index in the direction of propagation as sinusoidal, the effective refractive index can be written as
\begin{equation}
n(z)=n_\mathrm{eff}+\frac{\Delta n}{2} \cos \left(2 \beta_0 z\right).
\end{equation}
The periodic structure has the effect of coupling the forward and backward waves in the waveguide. The equation for the electrical field propagating with a wavelength $\lambda$ and free-space propagation constant $k_0$=$2\pi/\lambda$ when the time factor exp($i\omega$t) is suppressed and all transverse and lateral variations are ignored is
\begin{equation}
\frac{d^2 E}{d z^2}+\left[n(z) k_0\right]^2 E=0.
\end{equation}
Neglecting the term containing $\Delta n^2$, and using $\beta=n_\mathrm{eff}k_0$, we can write
\begin{equation}
\left[n(z) k_0\right]^2=\beta^2+4 \beta \kappa \cos \left(2 \beta_0 z\right),
\end{equation}
where $\kappa$ is the coupling strength (amount of reflection per unit length), given by
\begin{equation}
\kappa=\frac{\pi\Delta n}{2\lambda_B}.
\end{equation}
Now, in the case of a square-wave grating with duty-cycle $\mathrm{D}$, the effective refractive index perturbation can be written as
\begin{equation}
n_\mathrm{eff}(z)= \begin{cases} n_\mathrm{eff2} &   0 < z < \mathrm{D}\Lambda  \\ n_\mathrm{eff1} & \mathrm{D}\Lambda < z < (1-\mathrm{D})\Lambda \end{cases}
\end{equation}
The coupling strength for a $m^{th}$-order grating can be derived from the $m^{th}$-order Fourier component of the refractive index perturbation $\Delta n_\mathrm{eff}$ and is expressed as \cite{Wang_2013}
\begin{equation}
\kappa=\frac{\pi}{\Lambda} \frac{\Delta n_{\mathrm{eff}}}{\bar{n}_{\mathrm{eff}}} \frac{\sin (m \pi D)}{m \pi},
\end{equation}
where $\Delta n_{\mathrm{eff}}$=$n_{\mathrm{eff2}}$-$n_{\mathrm{eff1}}$.
Our \ce{Si3N4} gratings with circular posts cannot be described using a  sinusoidal or square-wave perturbation of refractive index. 
Therefore, to attain accurate values of coupling strength, we perform FEM simulations of our post-grating structure by sweeping the post-gap ($g_\mathrm{p}$) and the post diameter ($g_\mathrm{d}$) and mapping the variation of the effective refractive index in 2D space using interpolation. Then, using Fourier analysis, we calculate the coupling strength of our \ce{Si3N4} post-gratings structure as shown in Figure \ref{Fig:fft}. It can be seen in Figure \ref{Fig:fft}a (bottom) that a square-wave grating with duty-cycle 50\% has only odd harmonics and since our post-gratings exhibit the distribution of effective refractive index as shown in the top panel, it has a finite $\Delta n_{\mathrm{eff}}$ at even harmonics which determines the coupling strength of the Bragg grating. More accurate results could be obtained using a 3d simulation of the grating unit cell to determine the photonic bandgap \cite{joannopoulos1997photonic}, which is impractical for such weak grating structures due the requirement of a very fine simulation grid.

\noindent In order to determine the spectral response of the Bragg grating, we write the field as a sum of transmitted and reflected waves\cite{erdogan1997fiber}
\begin{equation}
E(z)=R(z) \exp \left(-j \beta_0 z\right)+S(z) \exp \left(j \beta_0 z\right).
\end{equation}
Here the functions $R(z)$ and $S(z)$ vary slowly as the rapidly varying phase factor are incorporated in the exponential functions. Also, we consider wavelengths that are near the Bragg wavelength $\lambda_B$ and therefore $\beta$=$\beta_0$+$\Delta \beta$ with $\Delta \beta << \beta_0$. Now, eq. (4) and eq. (8) are substituted into the wave eq. (3). Ignoring second derivative terms and collecting identical phase factor terms, we can arrive at the coupled-mode equations   
\begin{equation}
\begin{gathered}
\frac{d R}{d z}+j \Delta \beta R=-j \kappa S \\
\frac{d S}{d z}-j \Delta \beta S=j \kappa R,
\end{gathered}
\end{equation}
Using appropriate boundary conditions, we can solve the coupled-mode equations and write the solutions as
\begin{equation}
\left[\begin{array}{l}
R(z) \\
S(z)
\end{array}\right]=T\left[\begin{array}{l}
R(0) \\
S(0)
\end{array}\right]
\end{equation}
where $T=T_N * T_{N-1} * \ldots . T_i \ldots . T_1$. The grating can be divided into a number of sections $N$ and each section has a transfer matrix $T_i$ given by
\begin{equation}
T_i=\left[\begin{array}{cc}
\cosh \left(\gamma z_i\right)-i \frac{\Delta \beta}{\gamma} \sinh \left(\gamma z_i\right) & -\frac{\kappa}{\gamma} \sinh \left(\gamma z_i\right) \\
i \frac{\kappa}{\gamma} \sinh \left(\gamma z_i\right) & \cosh \left(\gamma z_i\right)+i \frac{\Delta \beta}{\gamma} \sinh \left(\gamma z_i\right)
\end{array}\right]
\end{equation}
Here $\gamma^2=\kappa^2-\Delta \beta^2$ and bandwidth between the first nulls around the main reflection peak is given as $\Delta\lambda=\frac{\lambda_B^2}{\pi n_g} \sqrt{\kappa^2+(\pi / L)^2}$. 
The peak reflectivity for a given coupling strength $\kappa$ and length of the grating $\mathrm{L_g}$ is given by $\mathrm{R}=\mathrm{tanh}^2(\kappa \mathrm{L_g})$.

\section{Performance comparison with state-of-the-art integrated lasers.}
\noindent We compare the key performance metrics of laser output power, intrinsic linewidth, wavelength/frequency tuning range, actuation bandwidth (tuning speed) and actuation mechanism of state-of-the-art integrated lasers based on heterogenous/hybrid III-V semiconductors,  with a commercial fiber laser and deployed iTLA (integrated tunable laser assembly).
Our demonstrated laser shows state-of-the-art performance for external cavity and self-injection locked lasers, which approaches the fiber-laser coherence and enables high-speed frequency-agility by using monolithically integrated piezoelectric actuators.
Our lasers achieve a performance on par with the state-of-the-art heterogeneous/ hybrid III-V semiconductor-based lasers, and show greatly reduced fabrication complexity and cost.
This makes our demonstrated lasers suitable for applications not only in sensing but also in optical communications.
To provide more detailed in-depth performance comparisons, Table \ref{tab:comparison} summarizes the state-of-the-art prior works on integrated lasers based on integrated III-V group semiconductor lasers, in terms of the maximum demonstrated off-chip output optical power, the linewidth, tuning range and the tuning speed.

\begin{table}[htbp]
	\caption{\textbf{Comparison with reported photonic integrated lasers}  
	The major performance parameters of the demonstrated E-DBR laser are compared with selected integrated lasers based on integrated III-V group semiconductors.
	A typical commercially available fiber laser are also included for reference. n/a: not available. $\dagger$: on-chip power is provided while others are fiber-coupled output power.}
	\label{tab:comparison}
	\bgroup % some hacking to make the table less crowded, see: https://tex.stackexchange.com/questions/31672/column-and-row-padding-in-tables
	\def\arraystretch{1.2}
	\centering
	\begin{tabular}{lllccccc}
	\toprule
	\thead{Laser \\Platform} & \thead{Laser\\ Architecture} & \thead{Max. Power \\(mW)} & \thead{Intrinsic \\Linewidth (Hz)} & \thead{Tuning \\(nm/GHz)} & \thead{Actuation \\bandwidth (MHz)} & \thead{Actuation \\mechanism}  & Reference\\
	\hline
	\multicolumn{8}{c}{\textbf{Integrated III-V external-cavity lasers (ECL)}}\\ \hline
\ce{Si3N4}(hybrid)             & E-DBR (8 mm)     & $> 25$                          & $1.5\times 10^{3}$    & \thead{$>0.5$ nm \\$<1.5$ GHz}    & \thead{$<0.01$\\10}	& \thead{Thermo-optic\\Stress-optic}        &   \textbf{This work} \\
\ce{Si3N4}(heterogeneous)             & E-DBR (20 mm)        & $10^\dagger$                          & 400                    & n.a.    & n.a.	& n.a.        &   \cite{xiang_high-performance_2021} \\
\ce{Si3N4}(hybrid)             & E-DBR (20 mm)        & 12                          & 320                   & n.a.    & n.a.	& n.a.        &   \cite{xiang2019ultra} \\
Si(heterogeneous)       & E-DBR (15 mm)         & 37                        & $1\times 10^{3}$                        & n.a.        & n.a.    & n.a.	& \cite{huang2019sub} \\
\ce{Si3N4}(hybrid)         & Vernier         & 6             & 400                       & $<2$ GHz            & 10	& Stress-optic         & \cite{lihachev2023frequency}  \\%Yuyao Guo \textit{et al.}
\ce{Si3N4}(hybrid) & Vernier         & 23                          & 40                        & 70 nm       & $<0.01$		& Thermo-optic	& \cite{fan2020hybrid}  \\
\ce{LiNbO3}(hybrid)     & Vernier         & $5.5^\dagger$          & $11.3\times 10^{3}$       & \thead{20 nm \\$<2$ GHz}     & \thead{$<0.01$\\600}	& \thead{Thermo-optic\\Electro-optic}          & \cite{li2022integrated}  \\
Si(heterogeneous)       & Vernier         & 1.54                        & 95                        & 118 nm         & $<0.01$    & Thermo-optic	& \cite{morton2021integrated} \\%Paul A. Morton \textit{et al.} 
	\hline
\multicolumn{8}{c}{\textbf{Integrated III-V self-injection locked lasers (SIL)}}\\ \hline
\ce{Si3N4}(hybrid)             & E-DBR+cavity       & 10.5                          & 3.8                    & n.a.    & n.a.	& n.a.        &   \textbf{This work} \\
\ce{Si3N4}(heterogeneous)             & E-DBR+cavity        & n.a.                          & 3                    & n.a.    & n.a.	& n.a.        &   \cite{xiang_high-performance_2021} \\
\ce{Si3N4}(hybrid)              & Ring         & 1.5                          & 25                    & 2 GHz     & 10           & Stress-optic	&   \cite{lihachev_low-noise_2022}
\\
\ce{Si3N4}(hybrid)              & Ring         &    0.3                       & 0.04                    & n.a.     & n.a.           &  n.a.	& \cite{li_reaching_2021}% 0.01$^\ddagger$  
\\
\ce{LiNbO3}(hybrid)     & Ring         & 0.15         & $3\times 10^{3}$       & 600 MHz     & 10 MHz	& Electro-optic          & \cite{snigirev2023ultrafast}  \\

	\hline
	\multicolumn{8}{c}{\textbf{Commercial semiconductor lasers for optical communications}}\\ \hline
III-V       & DS-DBR              &40        & $<1\times10^5$                          & $> 40$ nm                  & n.a.    & n.a. &   \cite{lumentum_lumentum_nodate}
\\
	\hline
	\multicolumn{8}{c}{\textbf{Commercial low noise erbium doped fiber laser}}\\ \hline
	Fiber 	& Bragg grating              & $<25$          & $<100$        & 1 nm         & $< 0.02$     & Thermo-optic           & \cite{nkt}    \\%NKT Photonics 
	\hline
\end{tabular}
\egroup
\end{table}

\section{Pull-back fabrication process of electrodes}

\begin{figure*}[htb]
	\centering
	\includegraphics[width=\textwidth]{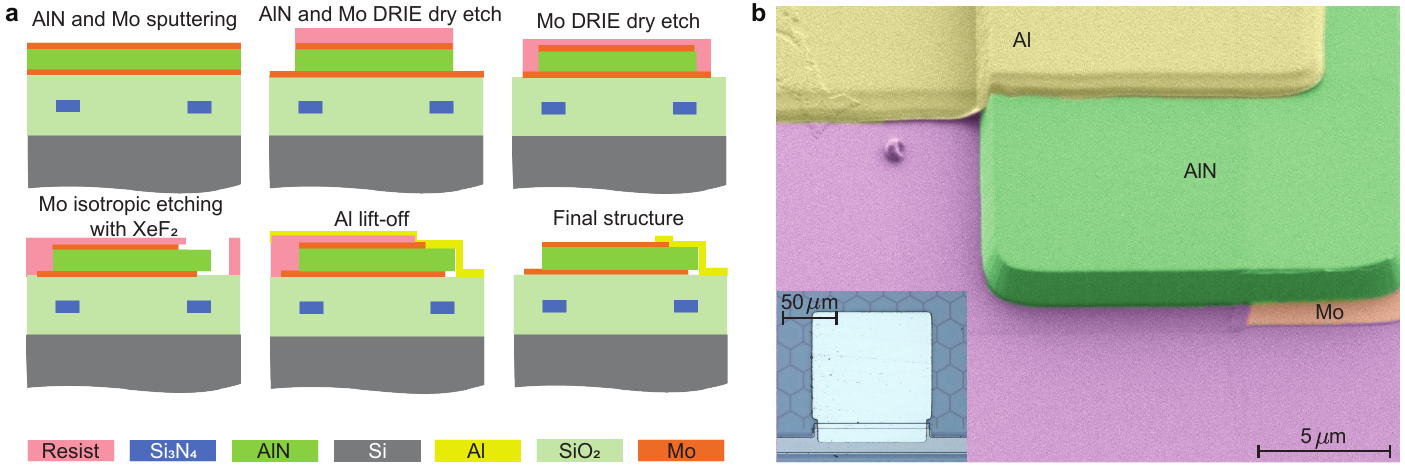}
	\caption{ 
		\footnotesize
		\textbf{Pull-back fabrication process.} (a) Process flow of the fabrication. (b) False-colored SEM image of the electrodes fabricated by this fabrication technique. The inset is a photograph taken by a optical microscope of the top electrode.
		}
	\label{Fig:Pull-back}
\end{figure*}

\noindent The "Pull-back process" for fabrication of electrodes is engineered to minimize the capacitance of bond-pads and for the ease of wire-bonding to the bond-pads.
As shown in Figure \ref{Fig:Pull-back}a, we start with a layer stack of top Mo, AlN, and bottom Mo on top of the SiO$_2$. 
We have used deep reactive-ion etching to pattern the Mo and AlN layers.
Prior to the pullback process, the top-electrode bond pad had a layer of AlN and bottom Mo underneath it, creating a capacitor.
Given the width of the bond pad is more than double the width of the device itself, the capacitance added is significant and should be avoided to reduce the RC time constant.
We can eliminate the capacitance by etching through the top Mo, AlN, and bottom Mo layers and depositing a layer of Al to connect the top electrode bond pad.
However, the deposited Al must be electrically isolated from the bottom Mo. 
Thus, we add the “Pull-back” step for which the process is named to create an air gap between the top and bottom metals.
XeF$_2$ etches Mo but has a near infinite selectivity with AlN. 
Therefore, using an isotropic etching process, the Mo can be slightly etched away without harming the AlN. 

\noindent We “Pull-back” the Mo from underneath the AlN between 1 and 2 $\mu$m, and then deposit a layer of Al with a lift-off process.
A glancing angle deposition tool evaporates 200~nm of Al with a 15$^\circ$ angle on a rotating plate to make sure the Al covers the 1 $\mu$m AlN step height.
The final result is a gap of air between the bottom Mo and top Al leaving them electrically isolated, but Al is electrically connected to the top Mo layer.
Figure \ref{Fig:Pull-back}(b) shows a trial of this technique, and it can be seen that the Mo has been pulled back from underneath the AlN and the Al has been deposited on top and fully covers the step height of the AlN.
In this trial, Al served as the entire top metal without top Mo, however the device described in this work has a layer of top Mo between the Al and AlN layers. 

%\FloatBarrier

\section{Bending loss and metal-induced propagation loss}

\begin{figure*}[htb]
	\centering
	\includegraphics[width=1\textwidth]{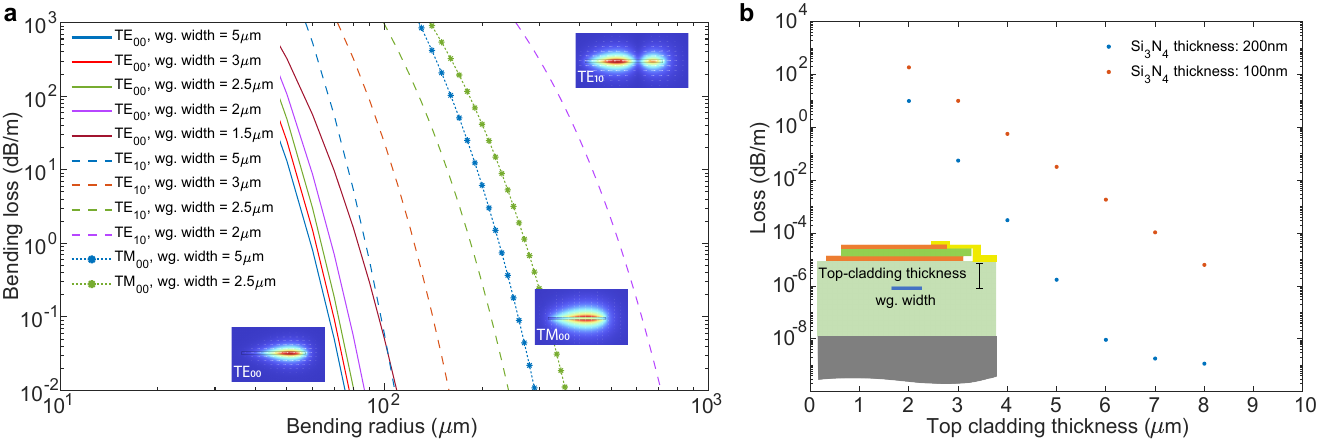}
	\caption{ 
		\footnotesize
		(a) Optical bending loss at different bending radii of the optical waveguide. The thickness of the \ce{Si3N4} core is kept constant at 200~nm. (b) Metal-induced propagation loss at different thickness of the top SiO$_2$ cladding.}

	\label{Fig:bending}
\end{figure*}

\noindent Figure \ref{Fig:bending}a shows the simulation of bending loss at different bending radii of the waveguide. The simulated results helped us to design photonic integrated circuits with high integration density. Also, it helped us to determine the cut-off bending radii for a particular width of the resonator to eliminate any excited higher-order TE or TM modes.
Figure \ref{Fig:bending}b shows the metal-induced propagation loss for different thickness of \ce{Si3N4} waveguide. The results show us that the top-cladding thickness of 3 $\mu$m for a 200~nm thick \ce{Si3N4} will provide negligible interaction between the optical mode and the metal electrodes of the piezoelectric actuator placed on top of the cladding. The metal will contribute only 0.1 dB/m propagation loss in this case. Whereas, if the thickness of \ce{Si3N4} is 100~nm, a 3 $\mu$m thick top cladding will not prevent metal interaction with the optical mode and the metal-induced propagation loss can be as high as 10 dB/m. In this case, a thicker SiO$_2$ top-cladding is needed. 

\section{Optical Frequency Domain Reflectometry (OFDR) measurement}

\begin{figure*}[htb]
	\centering
	\includegraphics[width=1\textwidth]{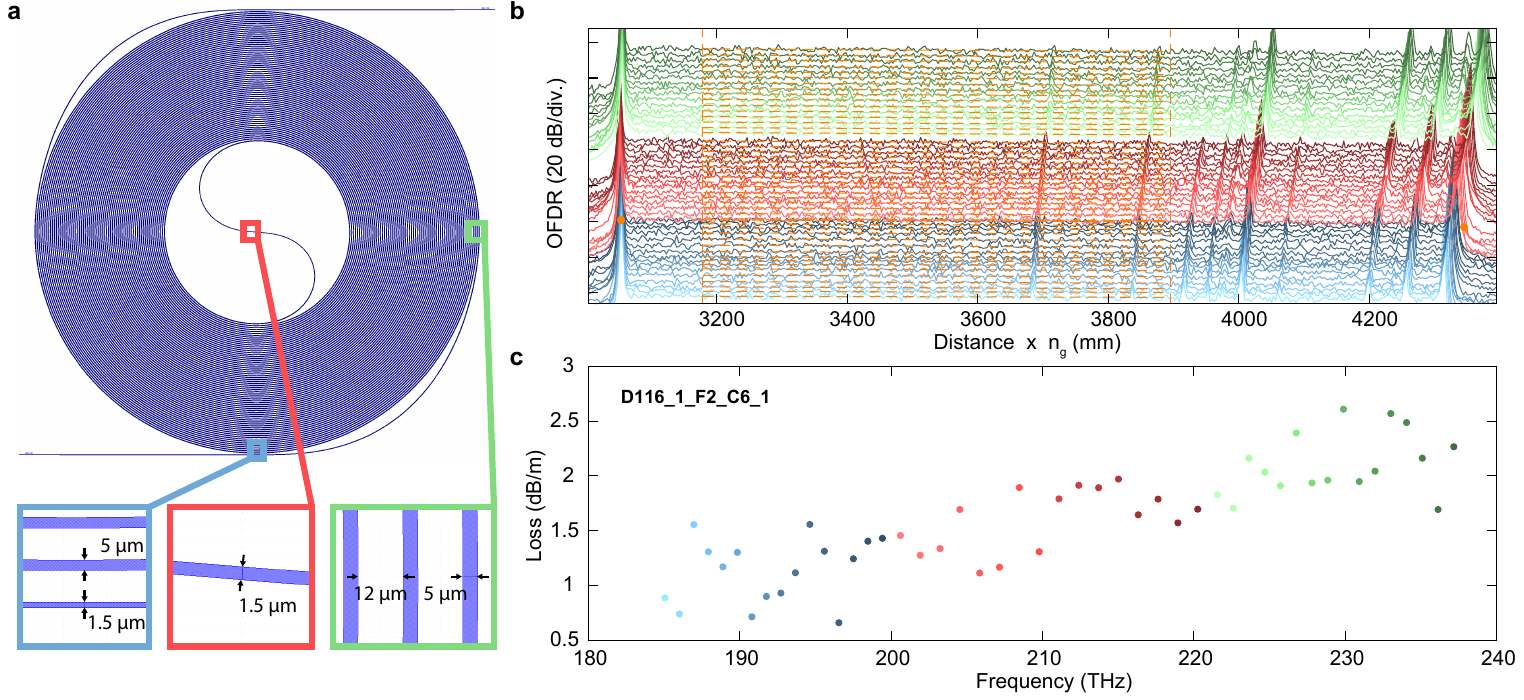}
	\caption{ 
		\footnotesize
		(a) GDSII layout of a 700 mm spiral waveguide with waveguide width 5~$\mu$m and single mode filtering sections of width 1.5~$\mu$m in the central S-bend (red inset) and the input section (blue and green inset). The waveguide width is tapered down to 1.5~$\mu$m at the centre of the spiral to eliminate higher order TE modes. (b) Optical frequency domain reflectrogram for central wavelength between 1260~nm and 1630~nm. Color coding similar to panel (c). (c) Propagation losses extracted from linear fitting of the reflection power decay in panel (b)}
	
	\label{Fig:ofdr}
\end{figure*}

\noindent We determine the optical propagation losses of the 200~nm thin \SiN-based waveguide platform via the analysis of long optical waveguide spirals. 
The spiral design is archimedian with C4-continuous spline bends that feature  smooth curvature and curvature derivative both at the straight waveguide interfaces and in the central S-bend\cite{chen2012general}. 
The waveguide width in the spiral arms are 5~$\mu$m wide and support one TM-like and three TE-like modes. 
To ensure single mode operation of the spiral and we apply a linear taper of the waveguide width in the central S-bend from 5 $\mu$m to 1.5 $\mu$m to strip higher order modes. 
We measure transmission, reflection, dispersion and propagation loss of a 70~cm long long 200~nm thin \SiN waveguide spirals using a custom-built, frequency-comb-calibrated scanning diode laser system\cite{del2009frequency}. 
Details of the optical setup are found in the supplementary materials of Ref.~\cite{riemensberger2022photonic}.
Three mode-hop-free, wideband-tunable, external-cavity diode lasers (ECDL,~Santec 710) are operated in sequence to form a full laser scan from 1260 to 1630~nm with calibration provided by a commercial fiber frequency comb (Menlo~FC1500-ULN).
Furthermore, an imbalanced fiber-optic Mach-Zehnder Interferometer (MZI) and a molecular gas cell are used for further calibration and wavelength determination.  
The transmission loss can be determined by Optical Frequency Domain Reflectometry (OFDR)\cite{lee2012ultra}.
The segmented Fourier transfrom of the OFDR interferogram of sample \texttt{D116\_1\_F2\_C8\_1} at center wavelengths between 1260~nm and 1630~nm is plotted in Figure~\ref{Fig:ofdr}(b).
Subsequent traces are offset by 3~dB. 
The front facet is found at an optical distance of 3050~mm, and the output facet is found at of 4300~mm. The strong wavelength dependence of the output facet indicates a strong normal dispersion of the thin waveguide spiral of 2000~fsmm$^{-1}$.
Sharp peaks indicate scattering centers in the spiral and are omitted in the subsequent fitting.
The propagation loss is determined by linear regression (orange) of the linear power decay of the backreflection and plotted in Figure~\ref{Fig:ofdr}(c). 
We find losses between 1~dB/cm at wavelength around 1600~nm rising to around 2~dB/cm at 1300~nm.
The frequency scaling of the losses is commensurate with Rayleigh-like scattering from sidewall imperfections being the dominant loss mechanism in our system. 
We do not detect a significant contribution from Si-H or N-H absorption, but cannot rule out contribution from other absorption channels such as metal impurities in the materials.

\section{RSOA-\ce{Si3N4} edge-coupling simulation}

\begin{figure*}[htb]
	\centering
	\includegraphics[width=1\textwidth]{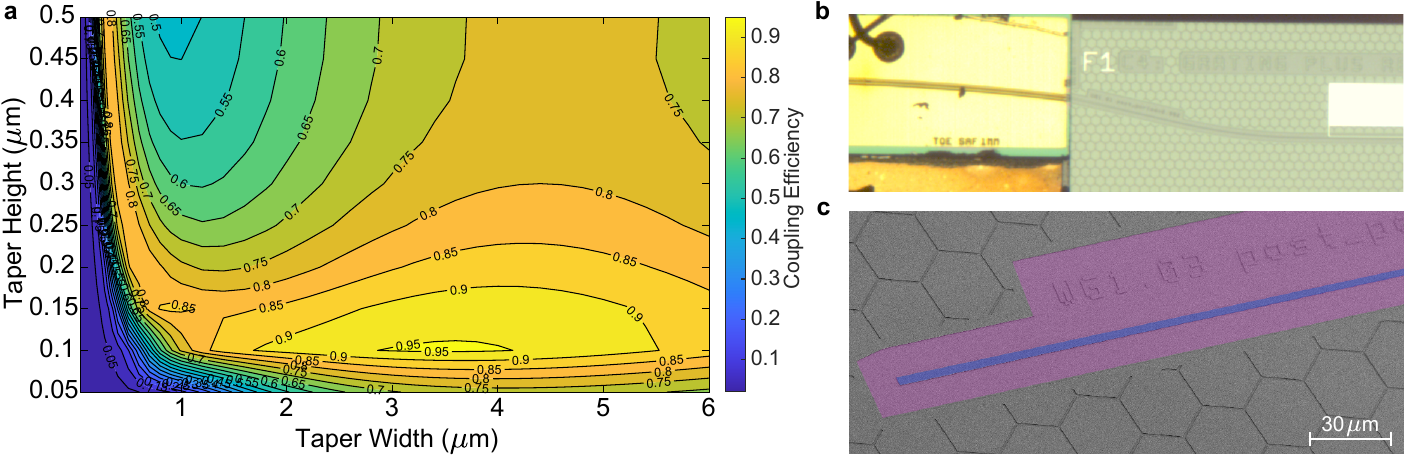}
	\caption{ 
		\footnotesize
		(a) Coupling efficiency estimation using overlap integral between the optical modes of the RSOA and the \ce{Si3N4} taper.  (b) Photograph if the edge-coupled RSOA and the \ce{Si3N4} PIC. (c) False-colored SEM of the horn-taper required for efficient coupling of light from the RSOA.}
	
	\label{Fig:taper}
\end{figure*}

Optical coupling between the \SiN waveguide and the RSOA is facilitated by waveguide tapers. We use so called horn tapers, i.e. the waveguide width from the chip to the edge is linearly increased from 2.5~$\mu$m to 4.5~$\mu$m in a distance of 300~$\mu$m. The angle of the taper is set to 12$^\circ$ to match the 6.8$^\circ$ waveguide angle and 19$^\circ$ degree free-space lateral beam exit angle of the RSOA according to Snell's law.
The coupling efficiency from the RSOA mode and \ce{Si3N4} taper mode can be estimated by mode overlap integral
\begin{equation}
    \eta = \frac{\left|\int_A E^*_\mathrm{RSOA}\cdot E_\mathrm{taper}dA\right|^2}{\int_A{\left|E_\mathrm{RSOA}\right|^2dA\cdot \int_A\left|E_\mathrm{taper}\right|^2dA}}, 
\end{equation}
where $E_\mathrm{\mathrm{RSOA}}$ is the mode field distribution at the RSOA facet and $E_\mathrm{taper}$ is the field distribution of the \SiN taper \cite{puckett2021broadband}.
Optical coupling between the \SiN waveguide and the RSOA is facilitated by waveguide tapers. We use so called horn-tapers, i.e. the waveguide width from the chip to the edge is linearly increased from 2.5~$\mu$m to 4.5~$\mu$m in a distance of 300~$\mu$m. The angle of the taper is set to 12$^\circ$ to match the 6.8$^\circ$ waveguide angle and 19$^\circ$ degree free-space lateral beam exit angle of the RSOA according to Snell's law.
Figure~\ref{Fig:taper}a depicts a numerical simulation of the overlap integral between the taper mode and the RSOA mode. The RSOA mode is inferred from the beam divergence angle. The best overlap for a thickness of 200~nm is found at for a waveguide width of 4.5~$\mu$m. We estimate the total coupling efficiency between the RSOA and the waveguide as 1~dB when accounting for reflection losses at the interface. In Figure \ref{Fig:taper}b, the photograph shows the butt-coupling of RSOA and the \ce{Si3N4} PIC and \ref{Fig:taper}c shows the SEM of the horn-taper widening from 2.5~$\mu$m to 4.5~$\mu$m over the length of 300~$\mu$m.
%%%%%%%%%%%%%%%%%%%%%%%%%%%%%%%%%%%%%%%%%%%%%%%%%%%%%%%%%%%%%%%%

\bibliographystyle{apsrev4-2}
\bibliography{zotero_updated}